\documentclass[journal]{IEEEtran}    
\usepackage{graphics}
\usepackage{graphicx}
\usepackage{bm}
\usepackage{indentfirst}
\usepackage{multirow}
\usepackage{cite}
\usepackage{url}
\usepackage[ruled,vlined]{algorithm2e}
\usepackage{makecell}
\usepackage{amsmath,amssymb,amsfonts}
\usepackage{booktabs}
\usepackage[margin=2cm]{geometry}
\usepackage{caption}

\usepackage{epsfig} 
\usepackage{amssymb}  
\usepackage{upgreek}
\usepackage{multirow}
\usepackage{float}
\usepackage[compatible]{algpseudocode} 
\usepackage[dvipsnames]{xcolor}
\usepackage{tikz}
\usepackage{lipsum}

\IEEEoverridecommandlockouts                              

\makeatother                                               

\title{\LARGE \bf
Practical Probabilistic Model-based Deep Reinforcement Learning by Integrating Dropout Uncertainty and Trajectory Sampling  
}

\author{Wenjun Huang$^{1,2}$, Yunduan Cui$^{2,*}$, Huiyun Li$^{2}$ and Xinyu Wu$^{2}$
\thanks{This research is supported in part by the National Natural Science Foundation of China under Grants 62103403; in part by Guangdong Basic and Applied Basic Research Foundation under Grant 2020B515130004.}
\thanks{$^{1}$ University of Chinese Academy of Sciences, China}
\thanks{$^{2}$ Guangdong-Hong Kong-Macao Joint Laboratory of Human-Machine Intelligence-Synergy Systems, Shenzhen Institute of Advanced Technology, Chinese Academy of Sciences, Shenzhen, China.}
\thanks{$^*$ Corresponding author: Yunduan Cui (e-mail: cuiyunduan@gmail.com)}
}

\begin{document}

\maketitle

\begin{abstract}

This paper addresses the prediction stability, prediction accuracy and control capability of the current probabilistic model-based reinforcement learning (MBRL) built on neural networks. A novel approach dropout-based probabilistic ensembles with trajectory sampling (DPETS) is proposed where the system uncertainty is stably predicted by combining the Monte-Carlo dropout and trajectory sampling in one framework.
Its loss function is designed to correct the fitting error of neural networks for more accurate prediction of probabilistic models. 
The state propagation in its policy is extended to filter the aleatoric uncertainty for superior control capability. 
Evaluated by several Mujoco benchmark control tasks under additional disturbances and one practical robot arm manipulation task, DPETS outperforms related MBRL approaches in both average return and convergence velocity while achieving superior performance than well-known model-free baselines with significant sample efficiency. The open source code of DPETS is available at \url{https://github.com/mrjun123/DPETS}. 
\end{abstract}

\section{Introduction}

Reinforcement learning (RL) provides a biomimetic learning framework where the agent gradually learns to complete the given task by interacting with the environment without prior human knowledge~\cite{sutton2018reinforcement}. 
As an appealing way of artificial intelligence, the model-free RL that directly learns control strategies without model knowledge has been widely applied to not only outperform humans in video and rule-based games~\cite{mnih2015human,silver2018general}, but also autonomously control complex systems including chemical plant~\cite{zhu2020scalable,dogru2022reinforcement}, unmanned vehicles~\cite{shi2018multi,li2019deep}, and robots~\cite{kober2013reinforcement,tsurumine2019deep}.
On the other hand, while model-free RL has achieved great success in simulation environments where sufficient and unbiased sampling is accessible, its real-world engineering applications remain challenging due to expensive sampling costs and complex environmental disturbances.
Model-based RL (MBRL) has been proposed to tackle these issues by approximating the system dynamics during the learning procedure. Although MBRL with a proper model contributes to superior sample efficiency compared to model-free RL, accurately modeling the system dynamics is difficult in engineering scenarios where uncertain disturbances lead to large modeling errors and can therefore cripple the effectiveness of MBRL.

One feasible solution is incorporating the system uncertainties in MBRL.
System uncertainties can be divided into two types: aleatoric uncertainty which arises from the inherent randomness of the system and epistemic uncertainty caused by a lack of system knowledge or data.
One famous probabilistic MBRL approach, the probabilistic inference for learning control (PILCO)~\cite{deisenroth2013gaussian} was proposed to employ Gaussian processes (GP)~\cite{rasmussen2006gaussian} and analytic moment-matching~\cite{girard2003gaussian,deisenroth2009analytic} to model and propagate the epistemic uncertainty of the target system in a full Bayesian perspective.
Although PILCO achieved great sample efficiency in many traditional control tasks, it assumes that all uncertainty can be fully observed and described by the system dynamics. This assumption is not suitable for engineering scenarios where environmental disturbances are usually unobservable and frequently changing.
To break this limitation, model predictive control (MPC) was employed in GP-MPC to promptly respond to the changing environment~\cite{kamthe2018data}. Its extensions have demonstrated great potential in unmanned surface vehicles in ocean environments~\cite{cuiJFR2020,cui2022filtered}.
However, implementing these approaches to high-dimensional systems with large samples remains difficult due to their non-parametric nature: the computational complexity of GP-based MBRL increases exponentially with the number of samples $N$, at a rate of $\mathcal{O}(N^{3})$.
Additionally, the aleatoric uncertainty in GP-based MBRL is modeled by homoscedastic Gaussian distributions without considering its variability.

\begin{figure*}[t]
\centering
\includegraphics[width=2.0\columnwidth]{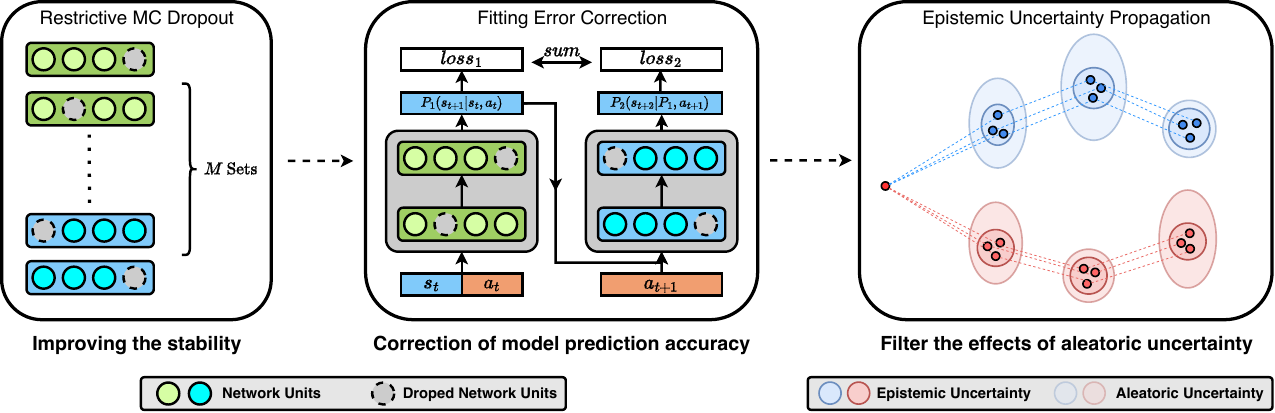}
\caption{Principle of the proposed DPETS in improving the stability of uncertainty propagation, correcting the fitting error, and filtering aleatoric uncertainty.}
\label{figure:1}
\end{figure*}

To address the issue of computational complexity described above, Deep Pilco~\cite{gal2016improving} was developed by employing deep neural networks, whose computational complexity does not depend on the number of samples, to approximate system dynamics. The epistemic uncertainty is estimated by a Monte-Carlo dropout (MC Dropout) with a theoretical guarantee, while the aleatoric uncertainty is also homoscedastically modeled. However, the prediction variance of MC Dropout highly depends on the dropout rate and neural network size, making it difficult to consistently express the uncertainty of the target system~\cite{osband2016risk}. 
Probabilistic ensembles with trajectory sampling (PETS) further introduced deep neural networks to MBRL with MPC-based policy~\cite{chua2018deep}. It employs a bootstrap sampling of trajectories to predict and propagate the epistemic uncertainty and utilizes multiple neural networks with Gaussian distribution output to estimate the aleatoric uncertainty.
Thanks to the robustness against disturbances under the MPC framework, PETS achieved comparable performances to model-free RL in many simulation benchmarks while enjoying superior sample efficiency.
Sharing the similar principle of PETS, model-based policy optimization (MBPO) further improved the sample efficiency in MBRL with long-term uncertainty propagation by generating enhanced data from the predictive model~\cite{janner2019trust}.
Overall, previous works~\cite{gal2016improving,chua2018deep,janner2019trust} have suggested using deep neural networks to overcome the computational burden of probabilistic MBRL and left the following issues unaddressed: 1) the uncertainty propagation based on neural networks is unstable; 2) the fitting error of neural networks is less considered; 3) the aleatoric and epistemic uncertainties are not distinguished during propagation.

In this paper, a novel and practical probabilistic MBRL approach, dropout-based probabilistic ensembles with trajectory sampling (DPETS)\footnote{Code available \url{https://github.com/mrjun123/DPETS}} was proposed to tackle all three issues above.
Following the principle of DPETS demonstrated in Fig.~\ref{figure:1}, the stability of uncertainty propagation was improved by introducing a restrictive MC Dropout in the trajectory sampling while the fitting error caused by neural networks was alleviated by a novel training strategy.
The aleatoric uncertainty with less Markov property was further filtered in propagation to properly estimate the epistemic uncertainty in the MPC-based policy.
Evaluated by several Mujoco benchmark control tasks with increasing complexity compared with related model-based and traditional model-free baselines, DPETS consistently demonstrated significant superiority in control performances, sample efficiency and robustness against disturbances.
In a practical robot arm control scenario, DPETS not only outperformed the model-free RL approaches while reducing $99\%$ usage of samples but also suppressed the related MBRL approaches with over $100\%$ higher average return and more sophisticated trajectories in robot end-effector control. All these results indicated the potential of DPETS as an emerging direction of practical MBRL approaches.

\begin {table}[t]
\caption{Comparison of the proposed and the related approaches}
\label{table:1}
\begin{center}
\resizebox{0.95\linewidth}{!}{
\begin{tabular}{|c|c|c|c|c|c|}
\hline
Approach  & MPC        & DNN       & \makecell{Fitting \\  Error}         & \makecell{ Distinguished \\ Uncertainties}  \\ \hline
PILCO~\cite{deisenroth2013gaussian}     & $\times$   & $\times$  & N/A                                                    & $\times$                                   \\ \hline
GP-MPC~\cite{kamthe2018data}    & $\bigcirc$ & $\times$  & N/A                                                   & $\times$                                   \\ \hline
Deep Pilco~\cite{gal2016improving} & $\times$   & $\bigcirc$& $\times$                                                        & $\times$                                   \\ \hline
PETS~\cite{chua2018deep}      & $\bigcirc$ & $\bigcirc$& $\times$                                                      & $\times$                                   \\ \hline
MBPO~\cite{janner2019trust}      & $\times$   & $\bigcirc$& $\times$                                                     & $\times$                                   \\ \hline
DPETS (ours)     & $\bigcirc$ & $\bigcirc$& $\bigcirc$                                            & $\bigcirc$                                \\ \hline
\end{tabular}}
\end{center}
\end{table}

According to the properties summarized in Table~\ref{table:1} where $\bigcirc$, $\times$ and N/A denote the involved, uninvolved and inapplicable terms, the contributions of this work are:
\begin{enumerate}
\item A restrictive MC Dropout with enhanced stability and expressive capability was proposed by combining the uncertainty propagation of Deep Pilco and PETS. It extended MC Dropout to more practical scenarios.
\item We explored a novel learning strategy to reduce the prediction bias caused by correcting fitting errors resulting from neural networks. This contributed to improved accuracy in multi-step prediction in probabilistic MBRL.
\item The uncertainty propagation of the MPC-based policy in MBRL was updated to filter out the aleatoric uncertainty that does not exhibit the Markov property. It effectively suppressed the rapidly expanding uncertainty in the multi-step prediction caused by environmental disturbances and therefore contributed to superior robustness and control capability against external disturbances.
\end{enumerate}

The remainder of this paper is organized as follows. Section \ref{S2} provided the background of existing probabilistic MBRL approaches based on neural networks. Section \ref{S3} detailed the proposed DPETS. It was evaluated and analyzed in Section~\ref{S4} as the experimental results. The conclusion was given in Sections \ref{S5}.

\section{Preliminary}\label{S2}
\subsection{Probabilistic MBRL with MPC-based Policy}\label{S2-1}

\begin{figure}[t]
\centering
\includegraphics[width=0.9\columnwidth]{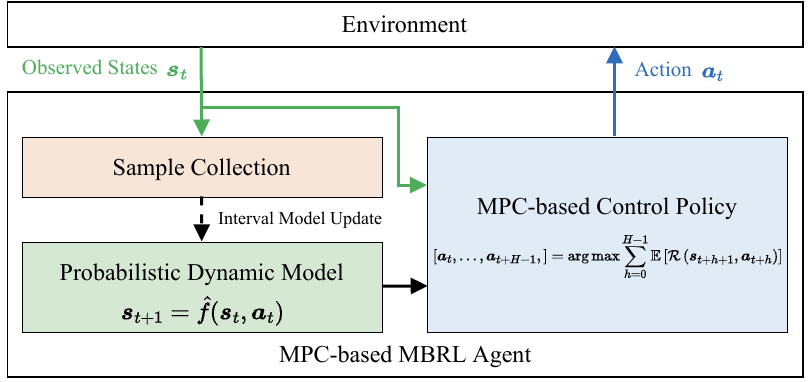}
\caption{Framework of probabilistic MBRL with an MPC-based policy.}
\label{figure:2}
\end{figure}

In this section, the probabilistic MBRL with an MPC-based policy is introduced following the demonstration in Fig.~\ref{figure:2}. 
The problem is modeled as a Markov decision process (MDP).
State and action at time step $t$ are defined as $\bm{s}_t\in \mathbb{R}^{d_s}$, $\bm{a}_t\in \mathbb{R}^{d_a}$.
The next step state follows the unknown system dynamics $\bm{s}_{t+1}=f(\bm{s}_t, \bm{a}_t)$.
Given a task-related reward function $\mathcal{R}(\bm{s}_{t+1}, \bm{a}_t)$, the agent aims to learn a policy $\pi:\bm{s}_t\rightarrow\bm{a}_{t}$ that maximizes the obtained reward in a long term.
Unlike model-free RL, which directly learns the value function of MDP, probability MBRL estimates the system dynamics from a Bayesian perspective:
\begin{gather} 
\begin{split}    
\left[\bm{\mu}(\bm{s}_{t+1}), \bm{\Sigma}(\bm{s}_{t+1})\right] = \hat{f}\left(\bm{s}_{t}, \bm{a}_{t}\right) + \bm{w}
\label{eq_dynamic_system}
\end{split}
\end{gather}
where $\bm{\mu}(\cdot)$ and $\bm{\Sigma}(\cdot)$ denote the mean and variance, $\bm{w}$ represents the aleatoric uncertainty caused by disturbances. 

GP-MPC~\cite{kamthe2018data} and PETS~\cite{chua2018deep} employ an MPC-based policy to promptly respond to frequently changing disturbances. Starting from the current state $\bm{s}_t$, the MPC-based policy predicts $H$ steps based on $\hat{f}(\cdot)$ and searches a sequence of optimal actions to maximize the long-term reward while propagating the uncertainty:
\begin{gather} 
\begin{split}    
\left[\bm{a}_t, \ldots, \bm{a}_{t+H-1}\right]=\arg \max \sum_{h=0}^{H-1} &\mathbb{E}\left[\mathcal{R}\left(\bm{s}_{t+h+1}, \bm{a}_{t+h}\right)\right],\\
\textrm{s.t.} \quad \left[\bm{\mu}(\bm{s}_{t+h+1}), \bm{\Sigma}(\bm{s}_{t+h+1})\right] &= \hat{f} \left(\bm{s}_{t+h}, \bm{a}_{t+h} \right).
\label{eq_mpc}
\end{split}
\end{gather}
It is optimized by nonlinear optimization approaches like the cross entropy method (CEM)~\cite{rubinstein2004cross}.
The MPC-based policy executes the first action $\bm{a}_t$ and moves to the next step, this process is repeated as a close loop controller $\pi:\bm{s}_t\rightarrow\bm{a}_{t}$ following Algorithm~\ref{alg:1}. 
The probabilistic model is updated by the collected samples $\mathcal{D}$ after each episode.

\subsection{Probabilistic Model with MC Dropout}\label{S2-2}

Deep Pilco~\cite{gal2016improving} built the probabilistic model by neural networks. It predicted the uncertainty by randomly sampling $Q$ sets of dropout particles following Bernoulli distribution $\left\{\bm{z}^{q}\right\}_{q=1}^Q,\bm{z}^{q}=\{\bm{z}^{q}_1, \ldots, \bm{z}^{q}_L\}$:
\begin{gather} 
\begin{split}    
\hat{\bm{y}_{t}}=\frac{1}{Q}\sum_{q=1}^{Q}\hat{f}_{\bm{W}}(\bm{x}_t,\bm{z}^{q})
\label{eq_nns}
\end{split}
\end{gather}
where $\hat{f}_{\bm{W}}(\cdot)$ was the neural networks with weights and bias matrix $\bm{W}$,  $\bm{x}_t = \{\bm{s}_t, \bm{a}_t\}$ was the input vector,  $\hat{\bm{y}}_t$ was the predicted state. Defining the first layer's input as $\hat{\bm{y}}_0=\bm{x}_{t}$, the output of layer $l=1, ..., L$ was calculated as:
\begin{gather} 
\begin{split}    
\hat{\bm{y}}_l=\phi\left(\left(\hat{\bm{y}}_{l-1} \circ \bm{z}_l^q\right) \bm{W}_l\right)
\label{eq_nns_output}
\end{split}
\end{gather}
where $\bm{W}_l$ is the weights matrix of the $l$-th layer, $\circ$ denotes the element wise product, $\phi(\cdot)$ is the activation function.

During the training, the loss function between ground truth $\bm{y}_{t}$ and the output $\hat{\bm{y}}_{t}$ was defined as the minus log likelihood over $Q$ sets of particles:
\begin{gather} 
\begin{split}    
L\left(\bm{x}_t, \bm{y}_{t}\right) =
\log\sum_{q=1}^{Q}
\exp\left(\frac{1}{2\sigma^2}\left\|\bm{y}
_{t}-\hat{f}_{\bm{W}}(\bm{x}_{t},\bm{z}^{q})\right\|^2\right)
\label{eq_nns_loss_2}
\end{split}
\end{gather}
where $\sigma$ is the manually selected homoscedastic standard deviation of the system noise $\bm{w}\sim \mathcal{N}(0,\sigma^2)$.

The uncertainty propagation was achieved by assuming the output as a Gaussian distribution. It can be treated as a sampling-based analytic moment-matching~\cite{girard2003gaussian}. 
Generate $P$ states from distribution $\bm{s}_{t}^{p} \sim \mathcal{N}(\bm{\mu}_{t}, \bm{\Sigma}_t)$, the output distribution was estimated as:
\begin{gather} 
\begin{split}    
\bm{\mu}(\bm{s}_{t+1})
&\approx \frac{1}{PQ}\sum_{p=1}^{P}\sum_{q=1}^{Q}\hat{f}_{\bm{W}}(\bm{s}^{p}_t, \bm{a}_t, \bm{z}^{p, q}),\\
\bm{\Sigma}(\bm{s}_{t+1})
&\approx \frac{1}{PQ}\sum_{p=1}^{P}\sum_{q=1}^{Q} \hat{f}_{\bm{W}}(\bm{s}^{p}_t, \bm{a}_t, \bm{z}^{p, q}){\hat{f}_{\bm{W}}(\bm{s}^{p}_t, \bm{a}_t, \bm{z}^{p, q})}^{T}\\&-\bm{\mu}_{t+1}\bm{\mu}_{t+1}^T + \sigma^2\bm{I}
\label{eq_mm}
\end{split}
\end{gather}
where $\bm{\Sigma}(\bm{s}_{t+1}^{p})$ mixtures both the aleatoric uncertainty from noise and the epistemic uncertainty from neural networks.

\begin{algorithm}[t]
    \SetKwFunction{RS}{Observe\_State}
    \SetKwFunction{MPC}{MPC\_Policy}
    \SetKwFunction{OA}{Execute\_Action}
        \KwIn{Sample set with warmup samples $\mathcal{D}$}
        Initialize the probabilistic model: $\hat{f}$\\
        \For{\rm episode $k=1, ..., K$}{
            \For{\rm time $t = 1, ..., T$}{
                $\bm{s}_{t}=$ \RS{}\\
                $\bm{a}_{t} =$ \MPC{$\bm{s}_{t}$} following Eq.~\eqref{eq_mpc}\\
                \OA{$\bm{a}_{t}$}\\
                $\bm{s}_{t+1}=$ \RS{}\\
                Expand sample set $\mathcal{D} \leftarrow \mathcal{D} \cup\left\{\bm{s}_t, \bm{a}_t, \bm{s}_{t+1}\right\}$.
            }
            Update $\hat{f}$ by sample set $\mathcal{D}$.
        }
      \caption{Framework of Probabilistic MBRL with MPC-based Policy}
    \label{alg:1}
     \end{algorithm}
     
\subsection{Uncertainty Propagation by Trajectory Sampling}\label{S2-3}

PETS provided an alternative solution to propagate uncertainties~\cite{chua2018deep}.
It directly approximated the aleatoric uncertainty as a distribution with mean and standard deviation:
\begin{gather} 
\begin{split}    
\left[\bm{\mu}_{\bm{W}}(\bm{x}_{t}), \bm{\Sigma}_{\bm{W}}(\bm{x}_{t})\right]=\hat{f}_{\bm{W}}(\bm{x}_{t}).
\label{eq_pets_nns}
\end{split}
\end{gather}
The epistemic uncertainty was represented through $B$ bootstrap ensambles of neural networks with independent weights $\{\bm{W}_1,\ldots,\bm{W}_B\}$:
\begin{gather} 
\begin{split}    
\hat{\bm{y}}_t&=\frac{1}{B}\sum_{b=1}^{B}\hat{f}_{\bm{W}_b}(\bm{x}_{t}), \\\hat{f}_{\bm{W}_b}(\bm{x}_{t})&\sim\mathcal{N}\left(\bm{\mu}_{\bm{W}_b}(\bm{x}_{t}), \bm{\Sigma}_{\bm{W}_b}(\bm{x}_{t})\right).
\label{eq_pets_predict}
\end{split}
\end{gather}
Define $\bm{E}_{\bm{W}_b}=[\bm{\mu}_{\bm{W}_b}(\bm{x}_{t})-\bm{y}_{t}]$, the loss function became:
\begin{gather} 
\begin{split}    
L\left(\bm{x}_t, \bm{y}_{t}\right)
\! &=\!\!\sum_{b=1}^B \!\bm{E}_{\bm{W}_b}^{T} \bm{\Sigma}_{\bm{W}_b}^{-1}(\bm{x}_{t})\bm{E}_{\bm{W}_b}\!\!+\!\log \operatorname{det} \bm{\Sigma}_{\bm{W}_b}\left(\bm{x}_{t}\right).
\label{eq_pets_loss}
\end{split}
\end{gather}

The aleatoric and epistemic uncertainties were not distinguished in PETS, each bootstrap ensemble employed $P$ independent trajectory sampling whose expectation contributed to the predicted state:
\begin{gather} 
\begin{split}    
\bm{\mu}(\bm{s}_{t+1})&=\frac{1}{BP}\sum_{b=1}^{B}\sum_{p=1}^{P}\bm{\mu}_{\bm{W}_b}(\bm{x}_{t}^{p}),\\
\bm{x}_t^{p}&\sim\mathcal{N}\left(\bm{\mu}_{\bm{W}_b}(\bm{x}_{t-1}^{p}), \bm{\Sigma}_{\bm{W}_b}(\bm{x}_{t-1}^{p})\right).
\label{eq_pets_predict_2}
\end{split}
\end{gather}

\section{Approach}\label{S3}

\begin{figure}[t]
\centering
\includegraphics[width=1.0\columnwidth]{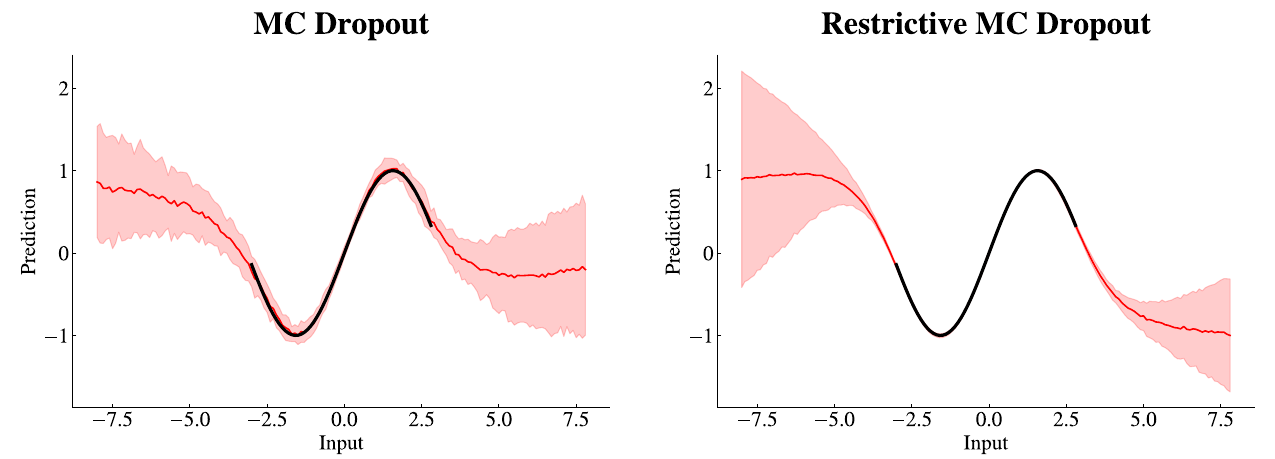}
\caption{Comparison of MC Dropout (left) and restrictive MC Dropout (right) in fitting noiseless sin functions. The training samples were shown in black line, the out epistemic uncertainty was demonstrated as the red region.}
\label{figure:3}
\end{figure}

\subsection{Restrictive MC Dropout}\label{S3-1}
The MC dropout proposed by Deep Pilco~\cite{gal2016improving} requires $Q$ times Bernoulli sampling for one prediction.
However, as shown in the left side of Fig.~\ref{figure:3}, 
its output uncertainty is highly dependent on the sampling results, which may unnecessarily increase epistemic uncertainty in the data-covered space and result in unstable control strategies.
To address this issue, DPETS proposes a restrictive MC Dropout with improved stability in uncertainty representation.
At the start of each episode, restrictive MC Dropout selects $M$ sets of dropout particles from a Bernoulli distribution $\left\{\bm{z}^{m}\right\}_{m=1}^M,\bm{z}^{m}=\left\{\bm{z}^{m}_1, \ldots, \bm{z}^{m}_L\right\}$.
During the episode, DPETS randomly samples $0.5M<Q<M$ sets of particles from the fixed $\left\{\bm{z}^{m}\right\}_{m=1}^M$. This contributes to a superior trade-off between presenting system uncertainties and avoiding the instability caused by unrestricted Bernoulli sampling.
Inspired by PETS~\cite{chua2018deep}, given the matrix of weights and bias $\bm{W}$ and the set of dropout particles $q$, the neural network simultaneously predict both mean and variance:
\begin{gather} 
\begin{split}    
\left[\bm{\mu}_{\bm{W}}\left(\bm{x}_t,\bm{z}^q\right), \bm{\Sigma}_{\bm{W}}\left(\bm{x}_t,\bm{z}^q\right)\right]=\hat{f}_{\bm{W}}\left(\bm{x}_t,\bm{z}^q\right).
\label{eq_RMC_nns}
\end{split}
\end{gather}

For a deterministic input $\bm{x}_t$, the output $\bm{y}_t$ is calculated by sampling $Q$ times with different sets of particles.
Compared to PETS, which models epistemic uncertainty using highly random sampling and described aleatoric uncertainty as homoscedastic Gaussian noise $\mathcal{N}(0, \sigma^2)$, 
DPETS effectively expresses both types of uncertainty while reducing unstable predictions, as shown in the right side of Fig.~\ref{figure:3}:
\begin{gather} 
\begin{split}    
\begin{aligned}
\hat{\bm{y}}_t & =\frac{1}{Q} \sum_{q=1}^Q\hat{f}_{\bm{W}}\left(\bm{x}_t, \bm{z}^q\right) \\
\hat{f}_{\bm{W}}\left(\bm{x}_t,\bm{z}^q\right) & \sim \mathcal{N}\big(\bm{\mu}_{\bm{W}}\left(\bm{x}_t,\bm{z}^q\right), \bm{\Sigma}_{\bm{W}}\left(\bm{x}_t,\bm{z}^q\right)\big).
\end{aligned}
\label{eq_RMC_nns_2}
\end{split}
\end{gather}
During the training, the loss function of neural networks using restrictive MC Dropout is defined as:
\begin{gather} 
\begin{split}    
L\!\left(\bm{x}_t, \bm{y}_t\right) \!\! =\!\!\sum_{q=1}^Q\! \!\left(\bm{E}_{\bm{W}}^T \bm{\Sigma}_{\bm{W}}^{-1}\left(\bm{x}_t, \bm{z}^q\right)\!\bm{E}_{\bm{W}}\!+\!\log\operatorname{det} \!\bm{\Sigma}_{\bm{W}}\!\left(\bm{x}_t, \bm{z}^q\right)\right)
\label{eq_RMC_loss}
\end{split}
\end{gather}
where $\bm{E}_{\bm{W}}=\left[\bm{\mu}_{\bm{W}}\left(\bm{x}_t,\bm{z}^q\right)-\bm{y}_t\right]$ is the difference between the output mean with particles set $q$ and the ground truth.


\subsection{Fitting Error Correction}\label{S3-2}
%

The loss functions of Deep Pilco and PETS only consider the error between the one-step prediction and the ground truth, as shown in Figs.~\eqref{eq_nns_loss_2} and \eqref{eq_pets_loss}. 
These loss functions do not distinguish between the errors caused by external disturbances and neural network approximation and therefore can easily lead to overfitting the one-step prediction that is heavily affected by system noises while neglecting the fitting error of dynamic models in a longer-term perspective.
This defect may result in accumulated bias and deteriorated control capability in the policies based on multiple-step predictions.
To tackle this issue, DPETS employs a novel loss function to further correct the fitting errors of neural networks by punishing incorrect predictions over two continuous steps.

Denote the sample of continuous two steps $t$ and $t+1$ as $\{\bm{s}_t, \bm{a}_t, \bm{s}_{t+1}, \bm{a}_{t+1}, \bm{s}_{t+1}\}$, $\bm{x}_t=\{\bm{s}_t,\bm{a}_t\}$, $\bm{y}_t=\{\bm{s}_{t+1}\}$, $\bm{y}_{t+1}=\{\bm{s}_{t+2}\}$,
the loss function with fitting error correction is calculated in two parts following the middle of Fig.~\ref{figure:1}:
\begin{gather} 
\begin{split} 
L_{FEC}(\bm{x}_t, \bm{a}_{t+1}, \bm{y}_t, \bm{y}_{t+1}) = L(\bm{x}_t, \bm{y}_t) \!+\! L'(\bm{x}_t, \bm{a}_{t+1}, \bm{y}_{t+1}).
\label{eq_RMC_loss_FEC}
\end{split}
\end{gather}
The first term focuses on the error of one-step prediction following Eq.~\eqref{eq_RMC_loss}. 
Define $\bm{x}_{t+1}^{q}=\{\bm{\mu}_{\bm{W}}(\bm{x}_t, \bm{z}^{q}), \bm{a}_{t+1}\}$ as the predicted mean in one-step prediction with particles set $q$,
the second term considers the corresponding error as:
\begin{gather} 
\begin{split} 
&L'(\bm{x}_t, \bm{a}_{t+1}, \bm{y}_{t+1})=\\&\sum_{q=1}^Q \left({\bm{E}'}_{\bm{W}}^T \bm{\Sigma}_{\bm{W}}^{-1}\left(\bm{x}_{t+1}^{q}, \bar{\bm{z}}^{q}\right){\bm{E}'}_{\bm{W}}+\log\operatorname{det} \!\bm{\Sigma}_{\bm{W}}\left(\bm{x}_{t+1}^{q}, \bar{\bm{z}}^{q}\right)\right)
\label{eq_RMC_loss_FEC_2}
\end{split}
\end{gather}
where ${\bm{E}'}_{\bm{W}}=\left[\bm{\mu}_{\bm{W}}\left(\bm{x}_{t+1}^{q},\bm{z}^q\right)-\bm{y}_{t+1}\right]$ is the difference between the output mean in two-step prediction with particles set $q$ and the ground truth $\bm{y}_{t+1}$, $\bar{\bm{z}}$ indicates another randomly selected set of particles independent to $\bm{z}$ for better generalization capability.
During the training process with $N$ samples, define $(\bm{x}^n, \bm{y}^n)$ as the $n$-th sample with states and actions over two continuous steps, the overall loss function of neural networks is set as:
\begin{gather} 
\begin{split} 
\mathcal{L}:=\frac{1}{N} \sum_{n=1}^N L_{FEC}(\bm{x}^n, \bm{y}^n) + \sum_{l=1}^L\lambda_l(\|\mathbf{W}_l\|_2^2+\|\mathbf{b}_l\|_2^2).
\label{eq_overall_loss}
\end{split}
\end{gather}
The second term punishes the over-large weights.

Following PETS~\cite{chua2018deep}, the proposed method utilizes $B$ ensembles of neural networks with independent matrix $\{\bm{W}_1, ..., \bm{W}_B\}$ to further enhance the model's representation capability of epistemic uncertainty besides the restrictive MC Dropout introduced in Section~\ref{S3-1}. 
Each ensemble of neural networks is trained with independent $Q$ randomly sampled sets of particles from the fixed $M$ sets in the current episode following Eq.~\eqref{eq_overall_loss}.

\subsection{Epistemic Uncertainty Propagation}\label{S3-3}

Existing works, Deep Pilco~\cite{gal2016improving} and PETS~\cite{chua2018deep} propagate their predictions via Eqs.~\eqref{eq_mm} and \eqref{eq_pets_predict_2} with consideration of both epistemic and aleatoric uncertainties. 
However, the aleatoric uncertainty caused by data noises and external disturbances which is independent of the system dynamics has a weaker Markov property, and merging it with the epistemic uncertainty of the system could result in over-large variances in multiple-step prediction which negatively affect the control performance of the policy.

We propose an efficient uncertainty propagation to filter the negative effect of aleatoric uncertainty in long-term prediction following the right side of Fig.~\ref{figure:1}.
In the $H$ step prediction of the MPC-based policy, each ensemble of neural networks with weights matrix $\bm{W}_b$ (we demonstrated the propagated states of two ensembles as blue and red color in Fig.~\ref{figure:1}) randomly selects $P$ sets of particles and conducts the following prediction:
\begin{gather} 
\begin{split}
\left[ \bm{\mu}_{\bm{W}_b}\left(\bm{x}_{t+h}^{b, p},\bm{z}^{b, p}\right), \bm{\Sigma}_{\bm{W}_b}\left(\bm{x}_{t+h}^{b, p},\bm{z}^{b, p}\right)\right] =\hat{f}_{\bm{W}_b}\left(\bm{x}_{t+h}^{b, p},\bm{z}^q\right)
\label{eq_propagation}
\end{split}
\end{gather}
where $\bm{x}_{t+h}^{b, p}=\{\bm{s}_{t+h}^{b,p}, \bm{a}_{t+h}\}$, $\bm{z}^{b, p}$ is the $p$-th set of particles selected by ensamble $b$, $h$ in the index of prediction horizon $H$.
The reward function in Eq.~\eqref{eq_mpc} at each step is calculated by the mean of all ensembles and particles to fully consider both epistemic and aleatoric uncertainties:
\begin{gather} 
\begin{split}
&\mathcal{R}(\bm{s}_{t+h+1}, \bm{a}_{t+h})\\&=\frac{1}{BP}\sum_{b=1}^{B}\sum_{p=1}^{P}\mathcal{R}\left(\bm{\mu}_{\bm{W}_b}(\bm{x}_{t+h}^{b, p},\bm{z}^{b, p}), \bm{\Sigma}_{\bm{W}_b}(\bm{x}_{t+h}^{b, p},\bm{z}^{b, p})\right).
\label{eq_reward}
\end{split}
\end{gather}
To filter the unnecessary aleatoric in long-term prediction, we omit the output variances in the propagation by setting the next step state as the output mean:
\begin{gather} 
\begin{split}
\bm{s}_{t+h+1}^{b,p} &= \bm{\mu}_{\bm{W}_b}(\bm{x}_{t+h}^{b, p},\mathbf{z}^{p}).
\label{eq_propagation_2}
\end{split}
\end{gather}
Only the epistemic uncertainty represented by $B$ ensembles and $P$ sets of particles would be passed to the next step. Please note that the uncertainty propagation introduced in this section is only employed in the MPC-based policy while the training process of $B$ ensembles is independently conducted through $Q$ sets of particles following Eq.~\eqref{eq_overall_loss} with full consideration of epistemic and aleatoric uncertainties.





\begin{algorithm}[t]
\SetKwFunction{RS}{Observe\_State}
\SetKwFunction{MPC}{MPC\_Policy}
\SetKwFunction{OA}{Execute\_Action}
\SetKwProg{Fn}{Function}{\string:}{end} 
    \KwIn{Sample set with warmup samples $\mathcal{D}$}
    Initialize ensambles of models $\hat{f}_{\bm{W}_b}, b = 1, ..., B$\\
    \For{\rm episode $k=1, ..., K$}{
        General $M$ fixed dropout particles $\left\{\mathbf{z}^{m}\right\}_{m=1}^M$ following Bernoulli distribution\\
        \For{\rm time $t = 1, ..., T$}{
            $\bm{s}_{t}=$ \RS{}\\
            $\bm{a}_{t} =$ \MPC{$\bm{s}_{t}$}\\
            \OA{$\bm{a}_{t}$}\\
            $\bm{s}_{t+1}=$ \RS{}\\
            \If{$t>0$}{Expand sample set $\mathcal{D} \leftarrow \mathcal{D} \cup\left\{\bm{s}_{t-1}, \bm{a}_{t-1}, \bm{s}_{t}, \bm{a}_t, \bm{s}_{t+1}\right\}$}
        }
        Update $\hat{f}_{\bm{W}_b}, b = 1, ..., B$ by sample set $\mathcal{D}$ based on $Q$ times sampling on $\left\{\mathbf{z}^{m}\right\}_{m=1}^M$ following Eqs.~\eqref{eq_RMC_loss_FEC}, \eqref{eq_RMC_loss_FEC_2} and \eqref{eq_overall_loss}
    }
    \smallskip
    \Fn{\MPC{$\boldsymbol{s}_{t}$}}{
        Sample dropout particles $\mathbf{z}^{b,p}$ from $\{\mathbf{z}^{m}\}_{m=1}^M$\\
        Set $B\times P$ initial states $\bm{s}_t^{b, p}=\bm{s}_t$\\
        Optimize sequence $[\bm{a}_{t}, ..., \bm{a}_{t+H-1}]$ by CEM to maximize 
        the reward in Eq. ~\eqref{eq_reward}, the states are propagated following Eqs.~\eqref{eq_propagation} and~\eqref{eq_propagation_2}
    }
    \Return{$\bm{a}_t$}
  \caption{Learning Procedure of DPETS}
\label{alg:2}
 \end{algorithm}

\begin{table*}[t]
\caption{Parameters of different control tasks}
\label{table:2}
    \resizebox{\linewidth}{!}{
        \begin{tabular}{|c|c|c|c|c|c|c|c|}
        \hline
        \textbf{Parameter}     & \textbf{Inverted Pendulum}     & \textbf{7-DOF Pusher}     & \textbf{HalfCheetah} & \textbf{Hopper}& \textbf{Ant} & \textbf{Walker2d} & \textbf{UrEEPosition}\\ \hline
        Layers number ($L$)  & 4    & 4    & 6  & 3 & 4 & 4 & 3\\ \hline
        Neuron number   &   200 & 200   & 200 & 200 & 200 & 200& 200\\ \hline
        Learning rate   & $10^{-3}$ & $10^{-3}$ & $10^{-3}$ & $10^{-3}$ & $10^{-3}$ & $10^{-3}$ & $10^{-3}$\\ \hline
        MPC horizon ($H$)   & 25    & 30    & 30   & 90 & 30 & 60 & 30 \\ \hline
        Rollout step ($T$)   & 200   & 150   & 1000  & 1000 & 1000 & 1000 & 300\\ \hline
        Dropout size ($M$)  & 5 & 5 & 5 & 5 & 5 & 5 & 5\\ \hline
        Ensambles ($B$)   & 5   & 5 & 5 & 5 & 5 & 5 & 5\\ \hline
        Parallel sampling ($P$)   & 4 & 4 & 4 & 5 & 4 & 4 & 4\\ \hline
        \end{tabular}
    }
\end{table*}

\subsection{Overview of DPETS}\label{S3-4}
In this section, we detail the whole process of the proposed DPETS following Algorithm~\ref{alg:2}.
Given the sample set $\mathcal{D}$ with warmup samples that are usually generated by random control actions, DPETS initializes $B$ ensembles of neural networks and interacts with the target environment in $K$ episodes.
At the start of each episode, $M$ fixed sets of particles are generated by restrictive MC Dropout. The agent then conducts a $T$ steps rollout. At each step, it observes the current state, decides and executes the control action, and observes the next step state. Unlike the general probabilistic MBRL framework, DPETS collects the samples over two continuous steps to fulfill its loss function with fitting error correction.
At the end of each episode, all $B$ ensembles of neural networks will be independently updated by $\mathcal{D}$ based on $Q$ times sampling on the fixed dropout particles following Eqs.~\eqref{eq_RMC_loss_FEC}, \eqref{eq_RMC_loss_FEC_2} and \eqref{eq_overall_loss}.

Based on the state propagation in Eqs,~\eqref{eq_propagation} and \eqref{eq_propagation_2}, the MPC-based policy in DPETS is conducted parallelly by $B$ ensembles of neural networks and $B\times P$ dropout particles.
Calculating the reward function by Eq. ~\eqref{eq_reward}, the optimization process of the control sequence $[\bm{a}_{t}, ..., \bm{a}_{t+H-1}]$ is achieved by CEM~\cite{rubinstein2004cross} following PETS~\cite{chua2018deep} to fairly evaluate the superiority of proposed method compared with the existing work.
Please note that it is straightforward to use other nonlinear optimization approaches in DPETS.
In practice, the output variance of DPETS is adaptively bounded following Appendix A.1 of PETS~\cite{chua2018deep}.

\begin{figure*}[t]
\centering
\includegraphics[width=2.0\columnwidth]{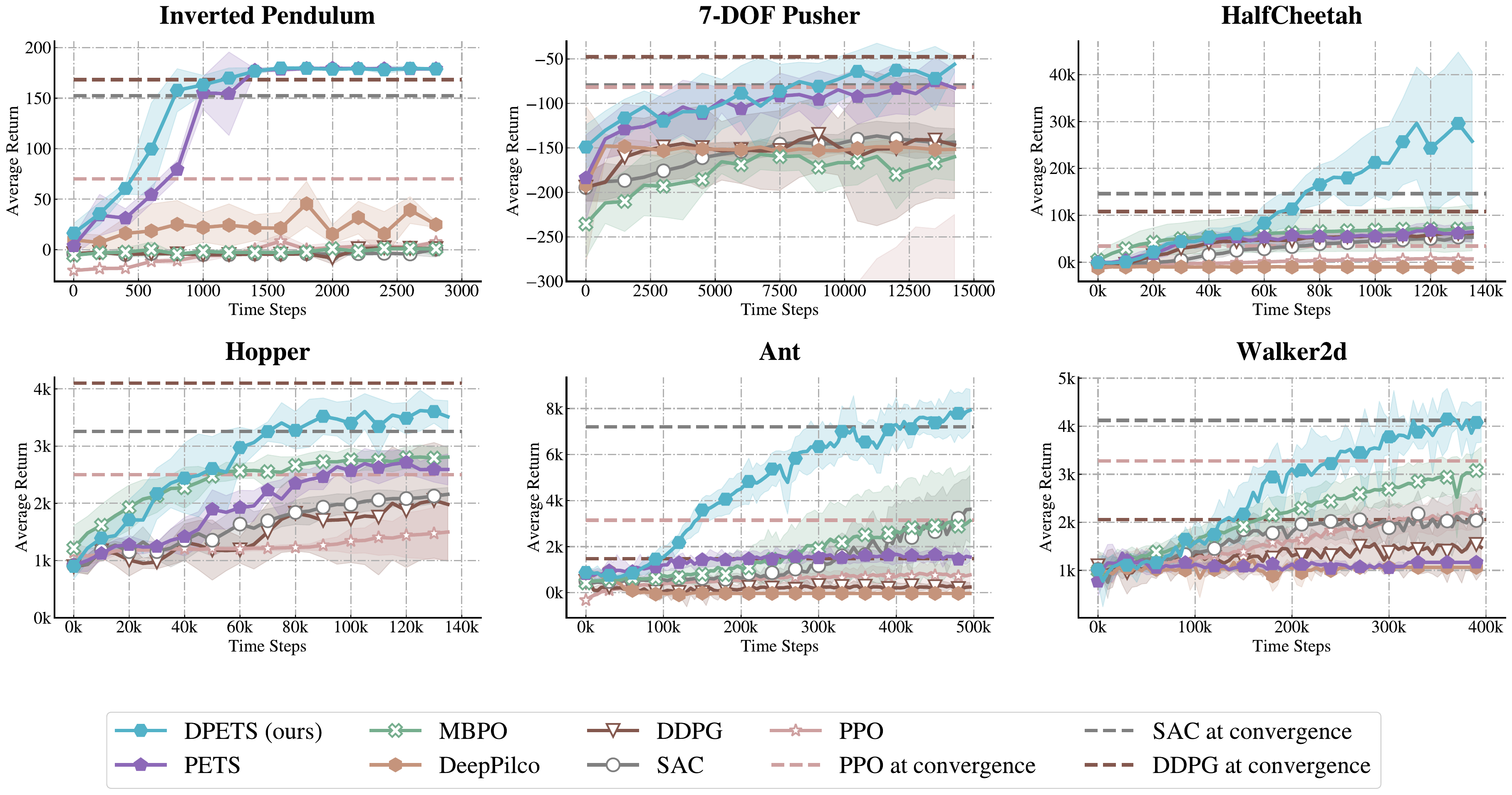}
\caption{Learning curves of DPETS and other baselines in Mujoco benchmark tasks. The shaded region represents the corresponding standard deviation.}
\label{figure:4}
\end{figure*}

\section{Experiment}\label{S4}
\subsection{Experimental Settings}\label{S4-1}
In this section, the proposed method DPETS was evaluated by six benchmark control tasks with increasing complexity: Inverted Pendulum, 7-DOF Pusher, Half Cheetah, Hopper, Ant, Walker2d and one practical robot arm end effector position control task (ur\_ee\_position). 
The proposed DPETS was implemented by PyTorch~\cite{NEURIPS2019_9015}.
We selected Deep Pilco~\cite{gal2016improving}, PETS~\cite{chua2018deep}, MBPO~\cite{janner2019trust} as the MBRL baselines, and selected SAC~\cite{haarnoja2018soft}, PPO~\cite{schulman2017proximal}, DDPG~\cite{lillicrap2016continuous} as the model-free RL baselines\footnote{Deep Pilco was developed following~\url{https://github.com/BrunoKM/deep-pilco-torch}, PETS and MBPO were developed based on~\url{https://github.com/kchua/handful-of-trials} and \url{https://github.com/jannerm/mbpo}. All model-free approaches were implemented by PARLe~\url{https://github.com/PaddlePaddle/PARL} and PaddlePaddl~\cite{ma2019paddlepaddle}.}.
Please note that the trick of multiple pieces of training in one episode in MBPO was disabled to ensure all MBRL methods are trained once per episode for a fair comparison.
The six benchmark tasks were developed based on OpanAI Gym~\cite{brockman2016openai} and Mujoco~\cite{todorov2012mujoco}. The practical robot manipulation task was from a robot-based simulator robogym~\cite{lucchi2020robo}.
They were conducted with the parameters summarized in Table~\ref{table:2}.
All experiments were conducted by three independent trials on a computational server with Intel i9-12900 CPU, NVIDIA GeForce RTX 3070 Ti GPU, 64GB memory and Ubuntu 18.04 OS.

\begin{figure*}[t]
\centering
\includegraphics[width=2.0\columnwidth]{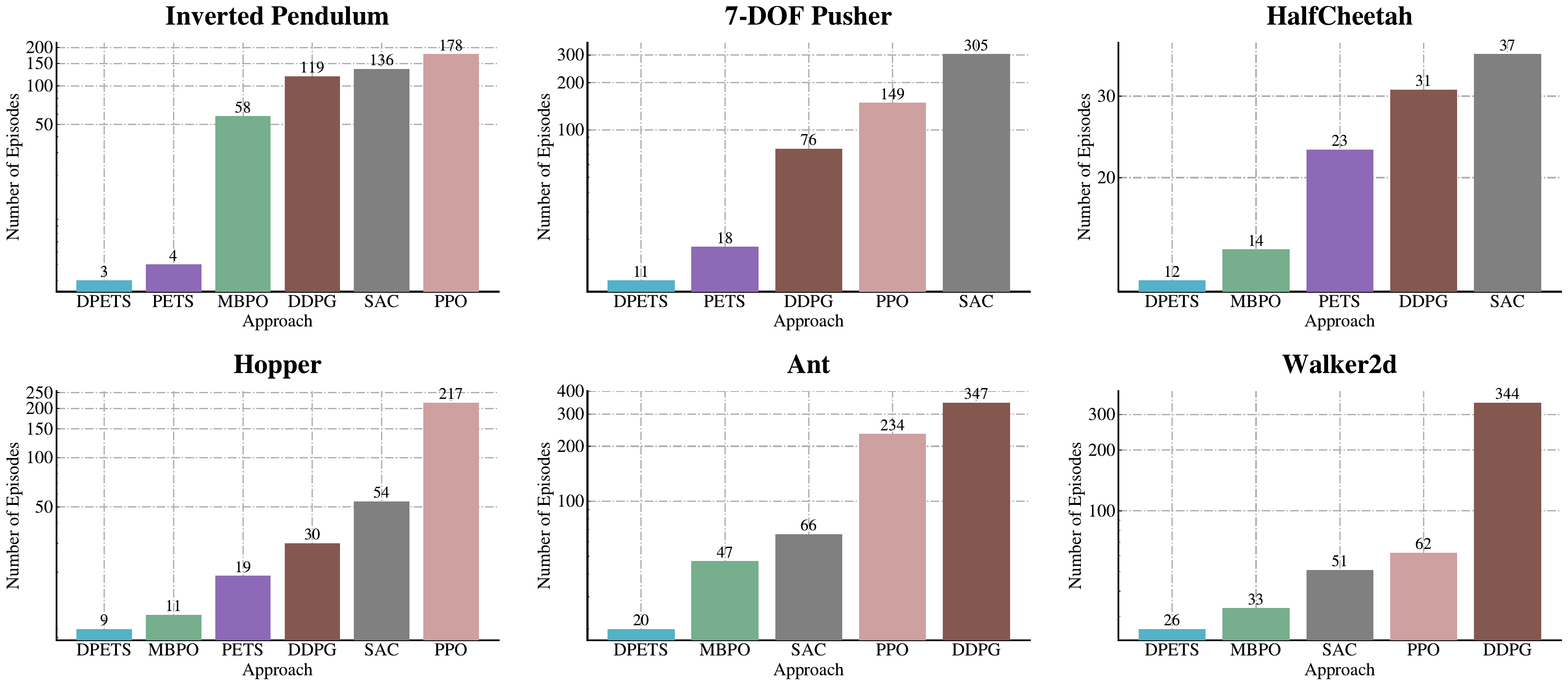}
\caption{Number of episodes used by all compared methods to reach the lower boundary of the maximum average returns over six benchmark control tasks.}
\label{figure:4_5}
\end{figure*}

\subsection{Evaluation of Control Performances}\label{S4-2}
We first evaluated the learning capability of the proposed method by six benchmark control tasks in Mujoco.
The learning curves DPETS and other baselines were compared in Fig.~\ref{figure:4}.
It was observed that DPETS enjoyed superiority in both average reward and convergence velocity. For other MBRL approaches, PETS had relatively good performances in the inverted pendulum and  7-DOF pusher tasks but converged slowly. MBPO worked in complicated scenarios like Hooper, Ant and Walker2d but could not reach the level of DPETS. Deep Pilco achieved the worst result and failed in all tasks except the inverted pendulum.

In the inverted pendulum task, DPETS significantly outperformed Deep Pilco and MBPO in the average reward through only $1.5k$ steps ($75$ episodes) interactions. PETS achieved a close performance with a slower convergence. Considering that all model-free approaches could not converge within $75$ episodes, we also compared their average reward after $200$ episodes (shown as dotted lines).
DPETS outperformed DDPG, SAC and PPO while reducing over $95\%$ interactions. This result indicated the excellent sample efficiency of DPETS.

In the 7-DOF pusher task, DPETS achieved a higher average reward than all MBRL baselines within $15k$ steps ($100$ episodes). Compared with the model-free baselines that converged with more than $150k$ steps ($1k$ episodes), DPETS outperformed SAC and PPO, while reaching a close performance to DDPG using only $10\%$ interactions. Please note that the standard deviation in the right bottom came from PPO which started from a very low average return near $-600$ and converged slowly.

In the HalfCheetah task, DPETS successfully outperformed all baselines within $140k$ steps. Compared with the suboptimal policy learned by SAC with $3000k$ steps, the proposed method achieved $180\%$ higher average reward with $95\%$ fewer interactions.

In the Hopper task, DPETS significantly outperformed other MBRL baselines within $70k$ steps (the result of Deep Pilco was not displayed due to its extremely low average return). Although DPETS achieved $18\%$ lower average return than DDPG that converged with more than $3000k$ steps (but still suppressed PPO at convergence and SAC at convergence with $36.2\%$ and $6.3\%$ higher average returns), it had significantly higher average return than all model-free approaches within the first $5\%$ interactions.

In the Ant task, DPETS quickly learned the best policy with about $8000$ average returns within $500k$ steps while other MBRL baselines' best performances were all below $4000$. Compared with the model-free RL baselines, it significantly outperformed the policies of DDPG, PPO and SAC that converged using six times more samples with over $437.1\%$, $133.6\%$ and $7.2\%$ higher average returns.

In the last Walker2d task, the proposed DPETS consistently demonstrated its advantages over all MBRL baselines within $400k$ steps, it outperformed MBPO, PETS and Deep Pilco with $19.8\%$, $277.7\%$ and $388.1\%$ more average returns. Compared with the best policy of SAC that converged after $3000k$ steps, DPETS achieved $92.4\%$ average return using only $13.3\%$ interactions. 

\begin{figure*}[t]
\centering
\includegraphics[width=2.0\columnwidth]{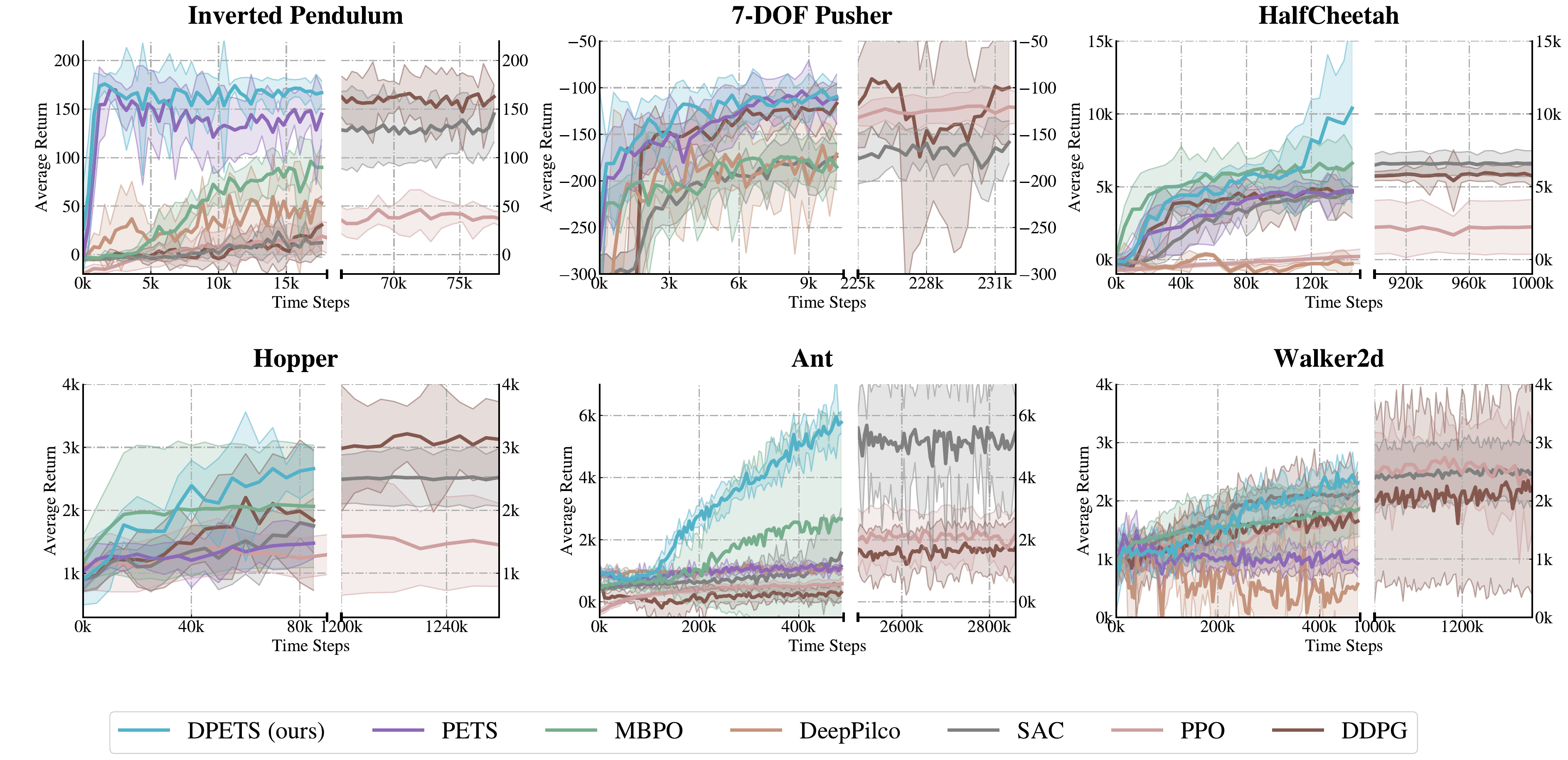}
\caption{Learning curves of DPETS and other baselines in Mujoco benchmark tasks with additional noises. The shaded region represents the corresponding standard deviation.}
\label{figure:5}
\end{figure*}

\subsection{Evaluation of Sample efficiency}\label{S4-3}

Defining the sample efficiency as the number of episodes spent by each method to reach the lower boundary of the maximum average returns over all baselines, we evaluated it in Fig.~\ref{figure:4_5} where the number of episodes was denoted by the first episode reaching the lower boundary based on the average learning curve over three random trials. Please note that the methods that could not learn the corresponding task were removed from the comparison in each subfigure.
In the inverted pendulum task, DPETS required only three episodes to reach the boundary while PETS and MBPO needed four and $58$ episodes respectively. In comparison, all model-free approaches converged to the same performance using over $100$ episodes.
In the 7-DOF pusher task, DPETS reduced the required episodes by $39\%$ and $85.5\%$ compared to PETS and DDPG. Although SAC finally reached a slightly superior average return, it required over $27$ times episodes to reach a certain level of performance.
In the HalfCheetah task, DPETS reached the lower boundary of maximum average returns while reducing the used episodes by about $14.3\%$ to $67.6\%$ compared with other baselines. 
In the Hopper task, the proposed method required only nine episodes to reach a certain level. Although DDPG achieved the highest maximum average return, it required about $40$ times as many interactions to meet the same level.
In the Ant task, only SAC had a close maximum average return to DPETS but had far worse sample efficiency. It required over three times as many samples to reach the lower boundary of the maximum average returns.
The same phenomenon was also observed in the Walker2d task, where DPETS converged to the lower boundary using only $78\%$ and $51\%$ samples of MBPO and SAC.
Overall, the proposed DPETS showed a significant advantage in sample efficiency compared to both model-based and model-free baselines across a wide range of benchmark control tasks. When given a certain level of control performance, DPETS converged to it with the fewest interactions, which demonstrates its potential in real-world hardware where sampling can be extremely expensive.

\begin{table*}[t]
\centering
\caption{Average maximum returns of DPETS with ablated components}
\label{table:3}
    \resizebox{1.0\linewidth}{!}{
        \begin{tabular}{|c|c|c|c|c|c|}
        \hline
        \textbf{Task}   & \textbf{DPETS}    & \textbf{DPETS-MC} & \textbf{DPETS-BE}  & \textbf{DPETS w/o FEC}   & \textbf{DPETS w/o DU} \\\hline
        \textbf{Pendulum} & {\color[HTML]{C73656}$\bm{179.29\pm1.26}$} & $176.65\pm 1.86$ & $175.72\pm1.21$ & $175.43\pm2.57$ & $176.16\pm2.19$\\ \hline
        \textbf{7-DOF Pusher} & $-62.76\pm7.43$ & $-84.66\pm8.37$ & $-79.52\pm25.9$ & $-86.65\pm 24.13$ & {\color[HTML]{C73656}$\bm{-61.35\pm8.82}$}\\ \hline
        \textbf{HalfCheetah} & {\color[HTML]{C73656}$\bm{25094.45\pm14872.13}$} & $7261.7\pm369.71$ & $21582.52\pm12059.06$ & $13926.72\pm5093.9$ & $9263.82\pm1025.2$\\ \hline
        \textbf{Hopper} & {\color[HTML]{C73656}$\bm{3427.88\pm252.01$}} & $1967.51\pm324.51$ & $2394.6\pm405.19$ & $1641.47\pm368.43$ & $1200.54\pm205.27$\\ \hline
        \textbf{Ant} & {\color[HTML]{C73656}$\bm{8420.22\pm554.37}$} & $929.83\pm4.79$ & $844.37\pm71.1$ & $931.62\pm72.35$ & $895.28\pm6.61$\\ \hline
        \textbf{Walk2d} & {\color[HTML]{C73656}$\bm{4151.32\pm242.39}$} & $3417.73\pm254.16$ & $3572.09\pm316.12$ & $2949.23\pm949.51$ & $2322.61\pm911.36$\\ \hline
        \textbf{Pendulum-N} & {\color[HTML]{C73656}$\bm{171.32\pm11.09}$} & $167.74\pm7.6$ & $155.4\pm15.14$ & $143.78\pm21.32$ & $152.99\pm18.51$\\ \hline
        \textbf{7-DOF Pusher-N} & {\color[HTML]{C73656}$\bm{-99.86\pm5.65}$} & $-102.94\pm12.67$ & $-110.96\pm15.34$ & $-106.16\pm15.34$ & $104.66\pm6.38$\\ \hline
        \textbf{HalfCheetah-N} & {\color[HTML]{C73656}$\bm{11320.19\pm2319.03}$} & $5831.79\pm853.59$ & $5610.18\pm609.79$ & $5408.26\pm297.97$ & $5576.09\pm159.98$\\ \hline
        \textbf{Hopper-N} & {\color[HTML]{C73656}$\bm{2434.6\pm 216.6}$} & $1380.45\pm104.52$ & $2125.39\pm587.48$ & $1308.2\pm125.4$ & $1326.22\pm287.8$\\ \hline
        \textbf{Ant-N} & {\color[HTML]{C73656}$\bm{5288.47\pm 382.37}$} & $911.38\pm124.52$ & $2126.32\pm188.47$ & $777.93\pm74.72$ & $867.1\pm158.78$\\ \hline
        \textbf{Walk2d-N} & {\color[HTML]{C73656}$\bm{2439.63\pm205.73}$} & $2084.54\pm221.54$ & $2379.27\pm164.34$ & $1702.43\pm106.89$ & $1302.44\pm145.32$\\ \hline
        \end{tabular}
    }
\end{table*}

\subsection{Evaluation of Noise Tolerance}\label{S4-4}

To further investigate the learning capability of DPETS against external disturbances, we added Gaussian noises with a factor of $0.05$ to the observed states in the six benchmark tasks above following~\cite{wang2019benchmarking}. According to the learning curves of average reward demonstrated in Fig.~\ref{figure:5}, all approaches required more samples and converged to a lower average reward under the additional disturbances while the proposed DPETS still enjoyed significant superiority.
In the inverted pendulum task, DPETS outperformed other MBRL baselines within $20k$ steps and reached the same performances of model-free baselines using only $25\%$ interactions.
In the 7-DOF pusher task, DPETS quickly converged to the height average reward compared with all baselines while the model-free approaches required $20$ times more samples to reach a close performance.
In the HalfCHeetah task, DPETS learned the best control policy within $150$ episodes, while the suboptimal policy learned by SAC required $1k$ episodes to reach $64\%$ average reward of DPETS.
In the Hopper task, DPETS was the only MBRL approach that quickly learned a comparable policy to the optimal one learned by DDPG after $1240k$ steps.
In the Ant task, DPETS quickly outperformed the suboptimal policy of SAC in the maximum average return using less than $20\%$ interactions. As a comparison, the suboptimal MBRL method MBPO reached the level of PPO with a very large standard deviation.
In the last Walker2d task, the proposed DPETS consistently had the best convergence and learning performances than other MBRL approaches. Using only $35\%$ number of interactions, it achieved a close control performance to both PPO and SAC while maintaining an acceptable range of standard deviation which indicated a more stable learning process.

\begin{figure}[t]
\centering
\includegraphics[width=1.0\columnwidth]{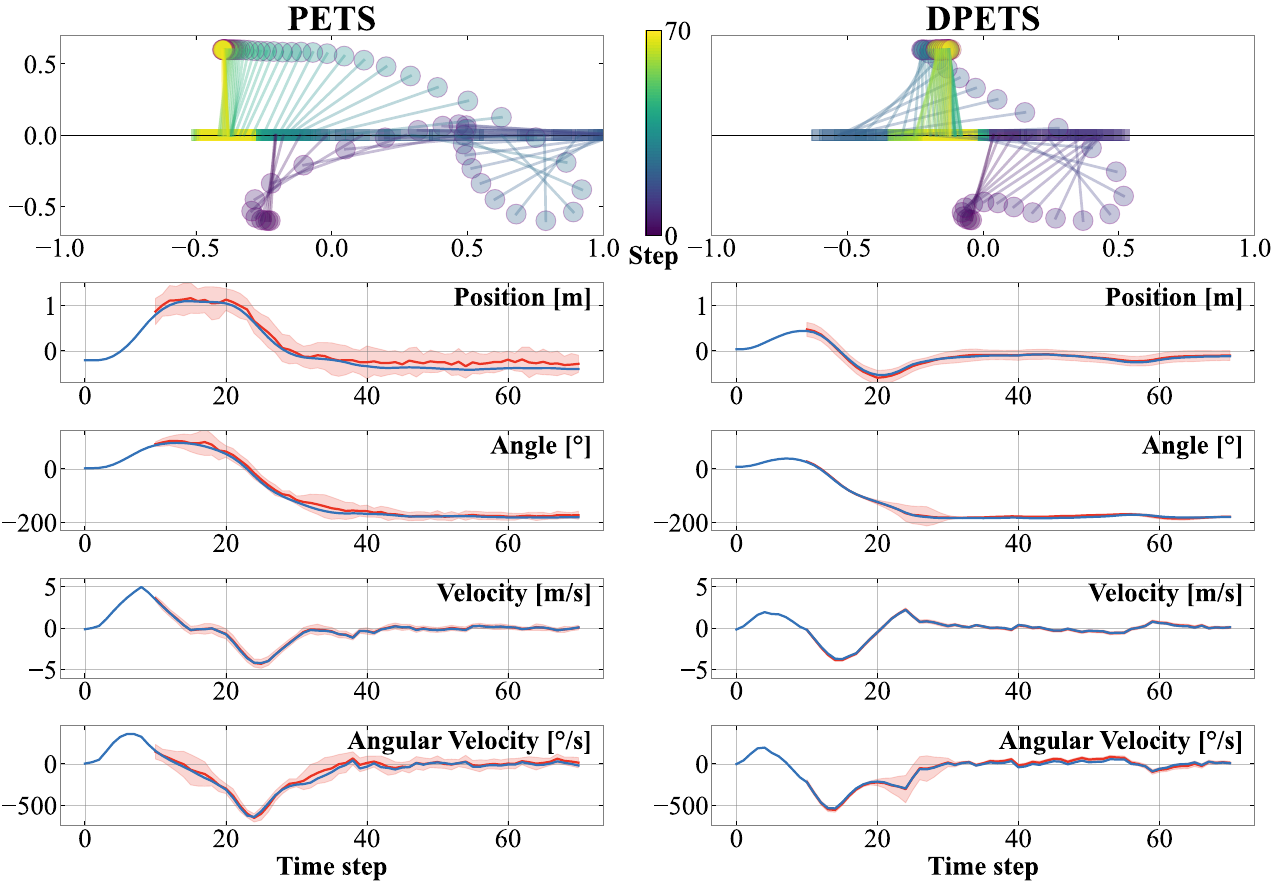}
\caption{Trajectories of states of PETS and DPETS in one test rollout of inverted pendulum with additional disturbances. The shaded region represents the predicted uncertainties.}
\label{figure:6}
\end{figure}

\subsection{Ablation Test}\label{S4-5}
The average reward of the learned policy in the ablation test was summarized in Table~\ref{table:3} where task with N indicates the additional Gaussian noises, DPETS-MC and DPETS-BE indicate the DPETS using the MC Dropout from Deep Pilco and the bootstrap ensembles from PETS, w/o FEC and DU indicate the proposed method without the loss function with fitting error correction and the propagation without distinguishing epistemic and aleatoric uncertainties.
These results demonstrated that the roles of each component (restrictive MC Dropout, fitting error correction and epistemic uncertainty propagation) in DPETS become increasingly important as task complexity and external disturbances increase.
In the original inverted pendulum task, three components had less impact on the converged policy.
Under Gaussian noises, MC Dropout resulted in $2\%$ fewer returns, bootstrap ensembles turned to $9\%$ fewer returns with a large standard deviation. Over $10\%$ fewer average reward was observed in DPETS w/o FEC and DU. 
In the 7-DOF pusher task, DPETS had $38\%$ fewer returns without using fitting error correction and epistemic uncertainty propagation. Under additional disturbances, DPETS achieved limited improvement ($2\%$ higher returns) which was consistent with the results of DPETS and PETS in Fig.~\ref{figure:5}.
In the original HalfCheetah task, DPETS-BE had $14\%$ fewer average returns while other ablated approaches could not learn the task with over $40\%$ reduced average returns. Turned to the task under noises, only DPETS successfully learned the task with over $50\%$ higher average returns than all ablated approaches.
In more complex scenarios such as Hopper, Ant, and Walker2d, fitting error correction and epistemic uncertainty propagation consistently contributed to a significant improvement in the maximum average returns, both in the original and noisy environments. Meanwhile, the MC Dropout and bootstrap ensembles resulted in deteriorated control performances.

\begin{figure}[t]
\centering
\includegraphics[width=1.0\columnwidth]{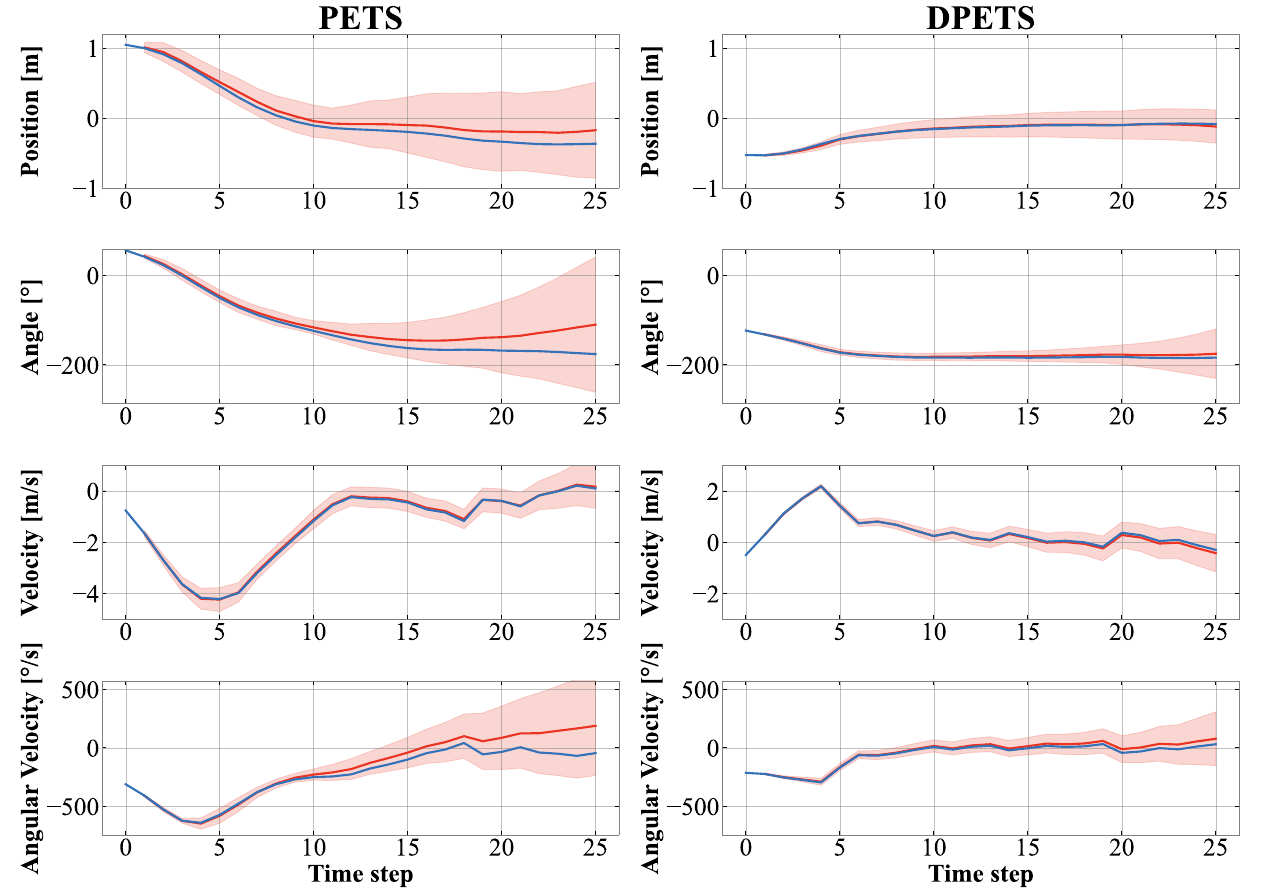}
\caption{Predicted trajectories of the MPC-based policy in PETS and DPETS at step $20$ in the rollout of inverted pendulum case study. The shaded region represents the predicted uncertainties.}
\label{figure:7}
\end{figure}

\begin{figure}[t]
\centering
\includegraphics[width=1.0\columnwidth]{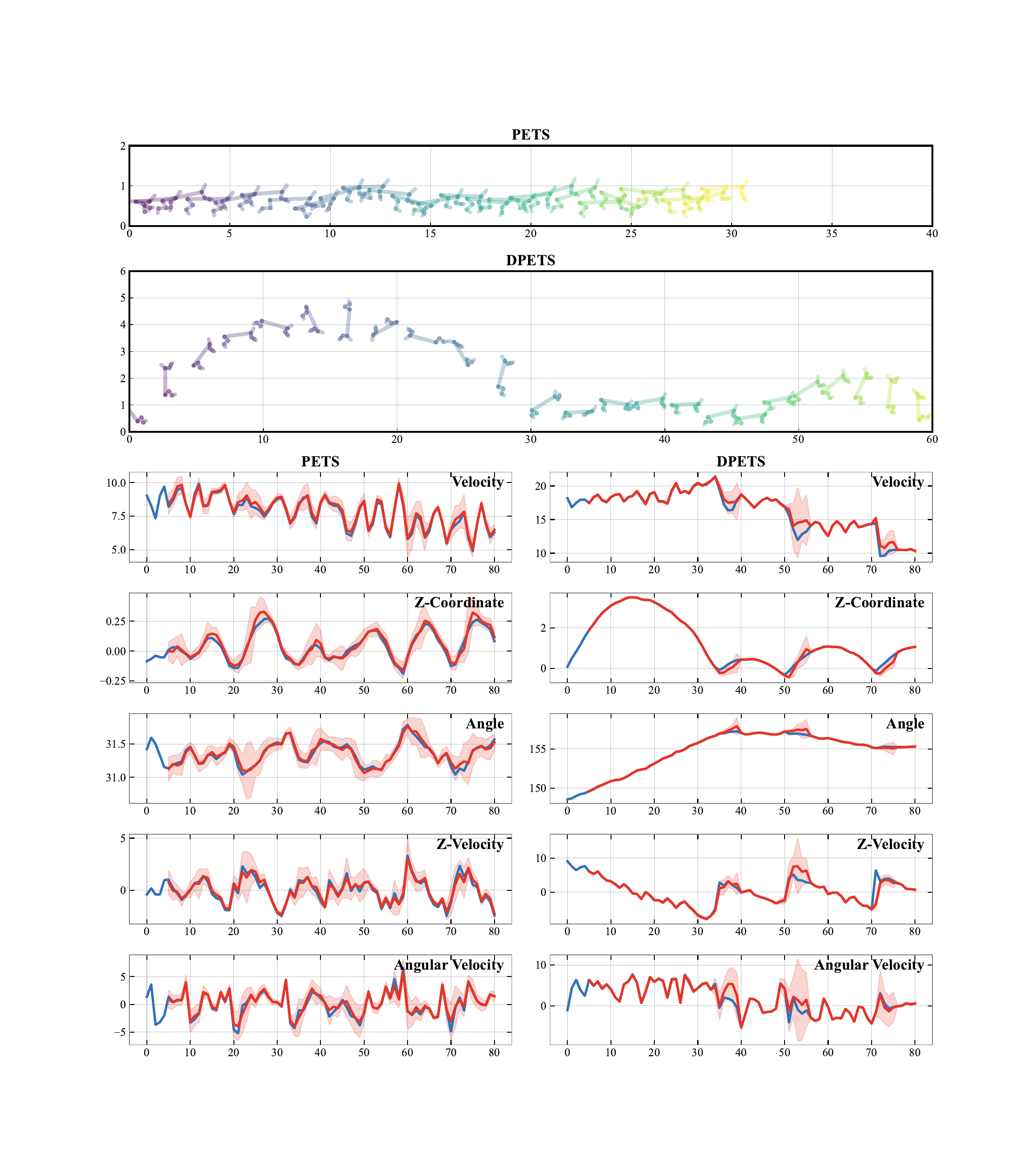}
\caption{Trajectories of states of PETS and DPETS in one test rollout of HalfCheetah with additional disturbances. The shaded region represents the predicted uncertainties.}
\label{figure:8}
\end{figure}

\subsection{Case Study}\label{S4-6}
In this subsection, we detailed one rollout of the policies learned by PETS and DPETS in the inverted pendulum task (after $100$ episodes) task and HalfCheetah (after $170$ episodes) task with disturbances as two case studies to further demonstrate the superiority of the proposed method compared with related MBRL approaches. 
The state trajectories in the inverted pendulum task were shown in Fig.~\ref{figure:6} where the blue lines are the trajectories of real strategies, the red lines are the trajectories of predicted states after $10$ steps, and the translucent areas represent the corresponding standard deviations.
It is observable that the proposed method enjoyed superior prediction accuracy than PETS in all dimensions.
Compared with the inaccurate prediction with a high standard deviation in PETS, the proposed method significantly alleviated the prediction error while properly describing the uncertainty by the standard deviation.
We analyzed the prediction of action sequences decided by the MPC-based policies of PETS and DPETS at step $20$ in Fig.~\ref{figure:7}.
Without the filter process of aleatoric uncertainty in the state propagation of the MPC-based policy, the predicted standard deviation of PETS rapidly increased over the full horizon and resulted in not only a hugely biased prediction but also an unreliable control sequence.
As a comparison, DPETS successfully removed the effect of aleatoric uncertainty and focused on the epistemic uncertainty generated by the model in prediction.
Based on the accurate estimation of future states, DPETS planned a more reliable control sequence to maximize the reward in the long horizon.

The superior characteristic of DPETS in control behavior and uncertainty propagation was also observed in the more challenging HalfCheetah task.
As shown in Fig.~\ref{figure:8}, unlike PETS which presented uncertainty even when the model estimation was accurate, the proposed DPETS only expressed high uncertainty in a few states where the estimation differed greatly from the observation and quickly turned to determined states. DPETS learned an efficient (but not very realistic) control strategy which is about twice as efficient in terms of movement per step compared with PETS. 
This result was consistent with the significantly higher maximum average return over all baselines in Fig.~\ref{figure:5}.
The two case studies above successfully indicated the superiority of DPETS. The probabilistic neural networks model using restrictive MC Dropout and loss function with fitting error correction enjoyed both improved accuracy in the long-term prediction and proper description of the aleatoric uncertainty. The MPC-based policy with epistemic uncertainty propagation greatly reduced the prediction error caused by the excessive inflation of aleatoric uncertainty in decision-making, significantly improving its control capability.

\begin{figure}[t]
\centering
\includegraphics[width=0.9\columnwidth]{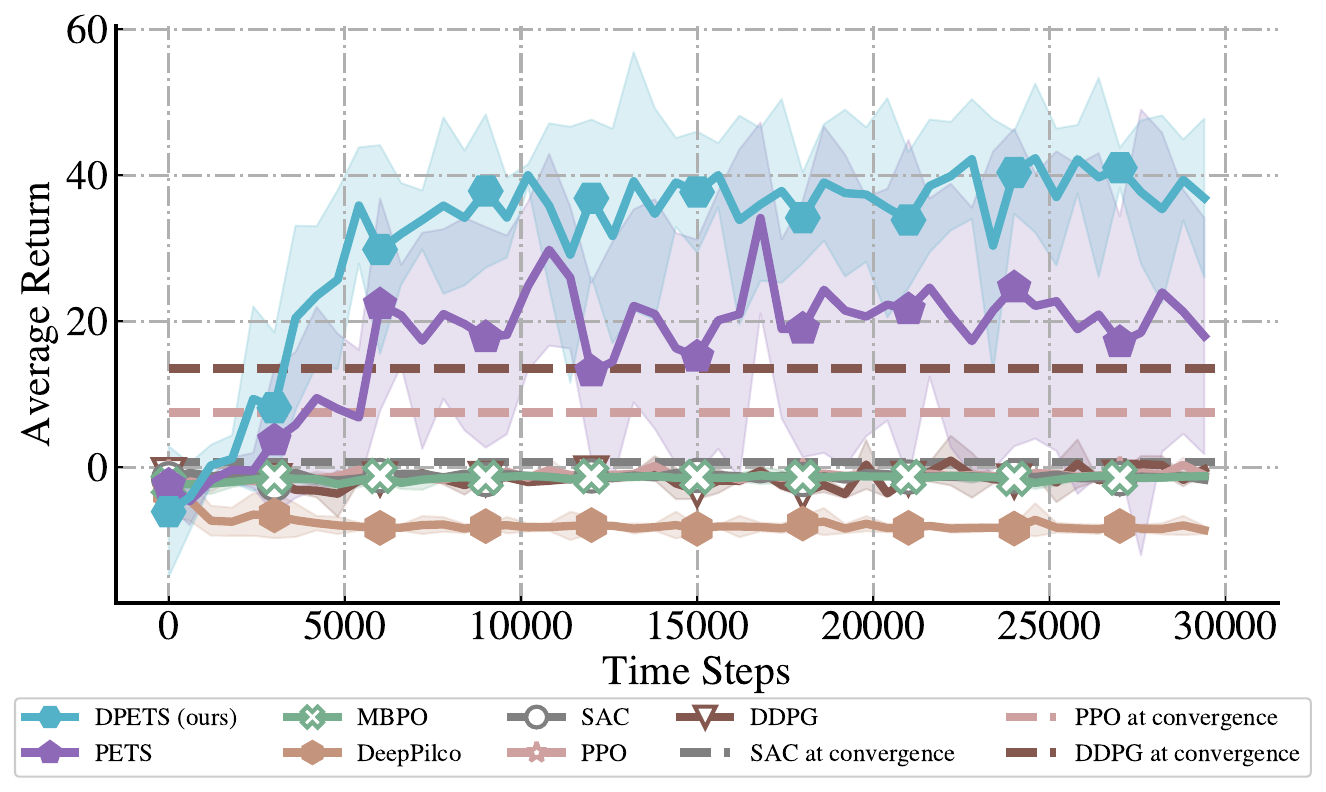}
\caption{Learning curves of DPETS and other baselines in UR5 end-effector position control task. The shaded region represents the corresponding standard deviation.}
\label{figure:9}
\end{figure}

\subsection{Robot Control Test}\label{S4-7}

Regarding the potential in engineering implementation, we finally evaluated DPETS by a real-robot-based simulation robogym~\cite{lucchi2020robo}. The target task ur\_ee\_position aims to control the UR5 robot arm to reach the randomly generated targets from its initial state. The learning curves of all compared methods were demonstrated in Fig.~\ref{figure:9}.
DPETS quickly converged to an optimal policy with about $40$ average return within the first $1000$ steps while PETS only reached about $20$ average return, neither MBPO nor Deep Pilco could learn this task.
As a comparison, all model-free baselines could not learn any meaningful policy in the first $30000$ steps. Even after $1000k$ interactions with the environment, none of them achieved a comparable performance to DPETS. Compared with the optimal model-free baseline DDPG, the proposed method learned a superior policy with over $100\%$ higher average return while reducing $99\%$ usage of samples. This result clearly indicated the great efficiency of DPETS in robot control.

\begin{figure}[t]
\centering
\includegraphics[width=0.9\columnwidth]{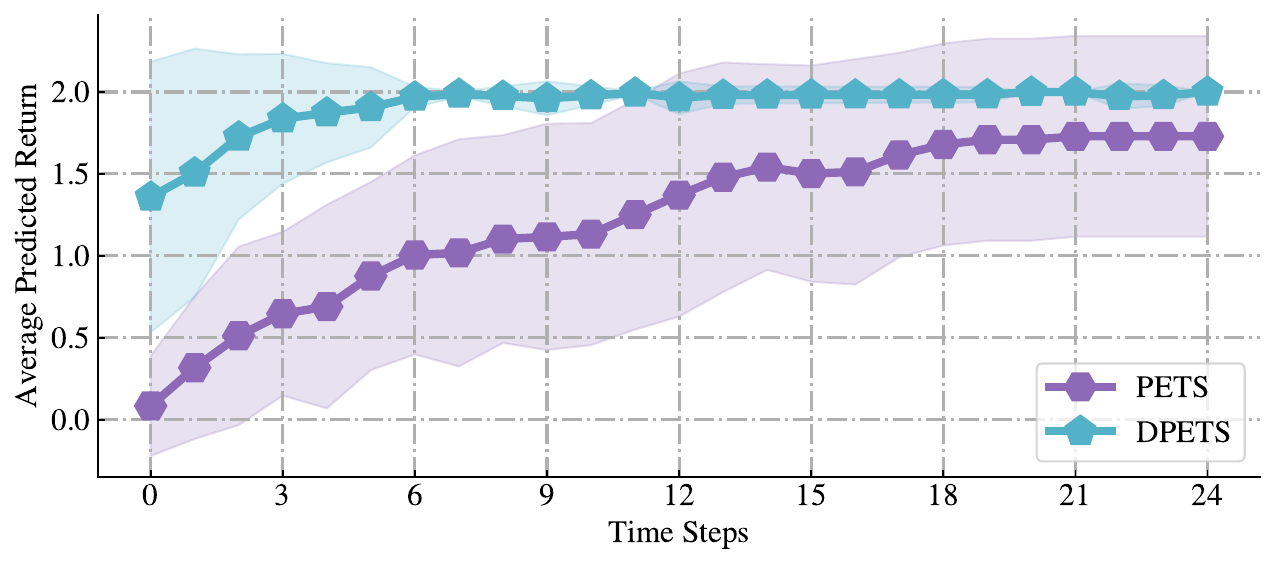}
\caption{The average predicted returns in the MPC prediction over $20$ evaluation rollout with $24$ steps. The shaded region represents the corresponding standard deviation.}
\label{figure:10}
\end{figure}

\begin{figure}[t]
\centering
\includegraphics[width=0.9\columnwidth]{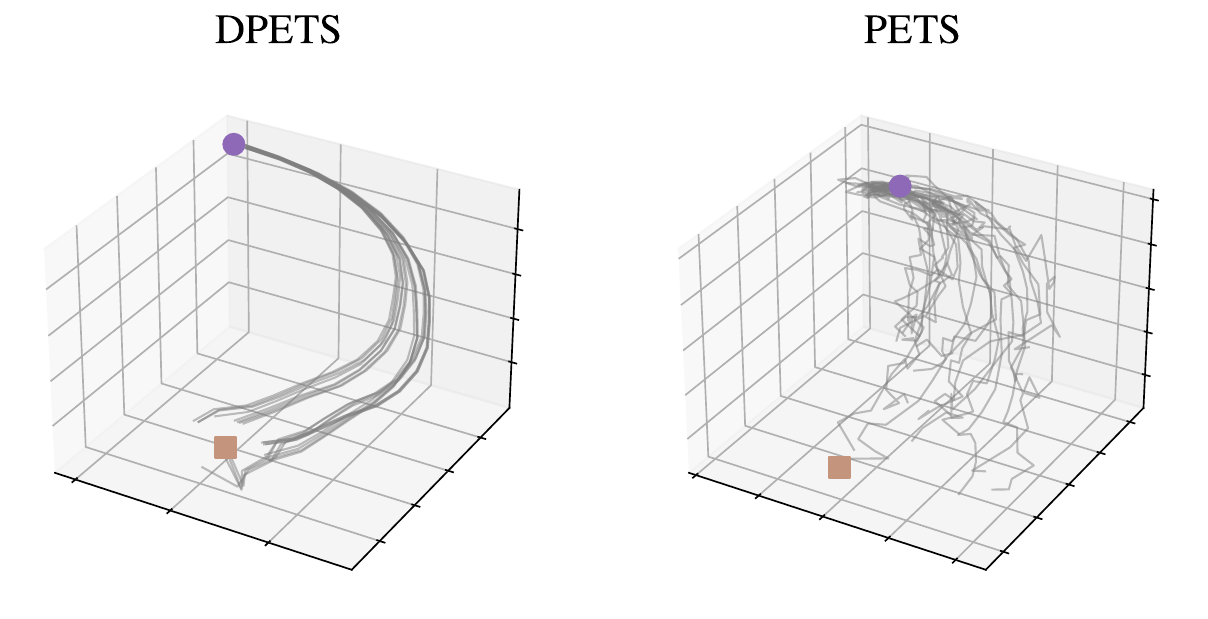}
\caption{The trajectories of robot end-effector optimized by DPETS and PETS.}
\label{figure:11}
\end{figure}

We further investigated the characteristics of DPETS in MPC and the corresponding behaviors. Evaluating the learned policies of DPETS and PETS after $30000$ steps' training, the average returns of the MPC prediction were summarized in Fig.~\ref{figure:10}. 
DPETS quickly optimized stable trajectories and converged the model uncertainty within $6$ steps and continuously executed the trajectories with confidence.
This feature contributed to smoother control trajectories of the robot arm in the space of the end-effector position during the MPC optimization as demonstrated in Fig.~\ref{figure:11}.
In contrast, PETS struggled to converge to confident control trajectories due to its unstable uncertainty propagation, resulting in not only lower average returns with overlarge predictive uncertainty but also frequently shaking control action of the robot arm.

\section{Conclusion}\label{S5}

In this paper, DPETS, a novel probabilistic MBRL approach based on neural networks was proposed to tackle the issues of prediction stability, prediction accuracy and policy quality in probabilistic MBRL. 
DPETS stably predicted the system uncertainty by introducing a restrictive MC Dropout that naturally combined dropout and trajectory sampling. 
A loss function with fitting error correction was proposed to reduce the approximation error of neural networks while improving its accuracy in long-term prediction.
An uncertain state propagation that filters aleatoric uncertainty was further developed to enhance the control capability of MPC-based policy.
Validated by six benchmark tasks under additional disturbances and one practical robot arm control task, DPETS not only outperformed the related MBRL approaches in average returns and convergence velocity but also achieved superior control performance compared with well-known model-free RL methods with significant sample efficiency.
These results indicated the potential of DPETS as a stable and sample-efficient MBRL approach to solve control problems under complicated disturbances from a probabilistic perspective.






\bibliographystyle{IEEEtran}
\bibliography{paper}

\end{document}